\documentclass[onecolumn,eqsecnum,aps,showkeys,showpacs,superscriptaddress, 11pt,prb]{revtex4-1}
\usepackage{epsfig}
\usepackage{dcolumn}
\usepackage{xspace}
\usepackage{color}
\usepackage[autostyle]{csquotes}
\usepackage{caption}
\usepackage[labelfont=bf,textfont=normalfont,singlelinecheck=off,justification=raggedright]{subcaption}
\usepackage{lineno,blindtext}

\begin{document}
\title{Role of {\color{black}{external and internal}} perturbations on ferromagnetic phase transitions in manganites: Existence of tricritical points}
\author{Prabir K. Mukherjee}
\affiliation{Department of Physics, Government College of Engineering
and Textile Technology, 12 William Carey Road, Serampore, Hooghly-712201, India}
\email{pkmuk1966@gmail.com}
\author{Prosenjit Sarkar}
\affiliation{
Department of Physics, Serampore College, Serampore 712201, India}
\email{psphysics1981@gmail.com}
\author{Amit K Chattopadhyay}
\affiliation{
Aston University, Mathematics, System Analytics Research Institute, Birmingham B4 7ET, UK}
\email{a.k.chattopadhyay@aston.ac.uk}
\date{\today}
\begin{abstract}
\centerline{Abstract} \noindent A phenomenological mean-field theory
is presented to describe the role of external magnetic field,
pressure and chemical substitution on the nature of ferromagnetic
(FM) to paramagnetic (PM) phase transition in manganites. The
application of external field (or pressure) shifts the transition,
leading to a field (or pressure) dependent phase boundary along
which a tricritical point is shown to exist where a first-order
FM-PM transition becomes second-order. We show that the effect of
chemical substitution on the FM transition is analogous to that of
external perturbations (magnetic field and pressure); this includes
the existence of a tricritical point at which the order of
transition changes. Our theoretical predictions satisfactorily
explain the nature of FM-PM transition, observed in several systems.
The modeling hypothesis has been critically verified from our
experimental data from a wide range of colossal magnetoresistive
manganite single crystals like Sm$_{0.52}$Sr$_{0.48}$MnO$_3$. The
theoretical model prediction of a tricritical point has been
validated in this experiment which provides a major ramification of
the strength of the model proposed.
\end{abstract}

\maketitle
\newpage

\section{Introduction}

The physics of critical phenomena and phase transitions is often a
study of pressure-temperature ambivalence defining the ubiquitous
phase plane that characterizes both first- and second-order phase
transitions. A classic example is the tricritical phase point where
all three phases simultaneously co-exist; while the discontinuous
jump of latent energy could drive first order phase transition,
\enquote{walking around} the phase boundary could be achieved
through continuous transition \cite{stanley,skma}. In most
ferromagnets, the transition from the high-temperature disordered
paramagnetic (PM) phase to the ferromagnetic (FM) ground state is
second-order and characterized by a continuous development of
magnetization below the transition point. But in some systems, FM
transition often demonstrates discontinuous change in magnetization
along the hysteresis path. Colossal magnetoresistive manganite is a
representative example of this class of systems. In manganites
RE$_{1-z}$AE$_z$MnO$_3$ (RE for rare earth ions and AE for alkaline
earth ions), the nature of phases and transitions strongly depend on
the bandwidth of the system as well as local disorder (also known as
quenched disorder), arising due to the size mismatch between RE and
AE cations
\cite{rao,dago1,dago2,toku1,hwang95,rod,QD-firstorder,tomi,sato}.
Such disorder reduces the carrier mobility, the formation energy of
lattice polarons which effectively truncates the FM phase and leads
the transition towards first-order \cite{sato}. It has been observed
that manganite with highest FM-PM transition temperature
($T_{FM-PM}$), La$_{1-z}$Sr$_z$MnO$_3$ with 0.2$<z<$0.5 undergoes
conventional second-order phase transition, whereas the lower
$T_{FM-PM}$ manganites such as Eu$_{1-z}$Sr$_z$MnO$_3$ with
0.38$<z<$0.47 show evidences of strong first-order FM transition
\cite{kghosh,esmo}. Although the order of phase transition is system
dependent, it can change under the influence of various parameters.
A change from first- to second-order FM transition
under the influence of various external and internal perturbations
is found in several theoretical
\cite{imry,atin,concen1,concen2,hysteresis,bustingorry} and
experimental works
\cite{kim,amaral,ultrasonic,mydeen,prb08,demko,prb09,prl09,sscmo,sr17,jmmm17}.
Among the manganite systems, particularly, which are very
susceptible to perturbations, Sm$_{1-z}$Sr$_{z}$MnO$_3$ (a narrow
band manganite with relatively large disorder) is one of the prime
candidate and a lot of analysis on FM-PM phase transition have been
done
\cite{mydeen,prb08,demko,prb09,prl09,isotope1,fisher,tomioka06,isotope2,apl1,apl2,murugeswari,ssmoru}.
For Sm$_{1-z}$Sr$_{z}$MnO$_3$ ($z$=0.45-0.48), FM transition at
ambient condition is first-order and the transition is extremely
sensitive to several parameters such as magnetic field, pressure,
chemical substitution or oxygen isotope exchange, etc.
\cite{mydeen,prb08,demko,prb09,prl09,isotope1,isotope2,apl1,apl2,murugeswari,ssmoru}.
In presence of external magnetic field ($H$) and pressure ($P$), the
FM transition shifts towards the higher temperature while the width
of thermal hysteresis in magnetization decreases gradually and
eventually vanishes at a critical magnetic field-pressure phase
point ($H_{C},\:P_{C}$), above which the transition becomes
second-order. Similar to external pressure, the application of
chemical/ internal pressure (which can be modulated by
stoichiometric control) also increases the $T_{FM-PM}$ as observed
in the partial substitution of Nd at Sm-site of
Sm$_{1-z}$Sr$_{z}$MnO$_3$
[(Sm$_{1-x}$Nd$_x$)$_{1-z}$Sr$_{z}$MnO$_3$] with $z$=0.45 and 0.48.
The effect of Nd doping on the nature of FM transition is quite
similar to that of external pressure, \emph{i.e.}, there exist a
critical concentration ($x_{C}$) above which the FM-PM phase
boundary changes from first to second order. In other words, using
parametric control of key pressure-temperature values, the first
order FM transition in Sm$_{1-z}$Sr$_{z}$MnO$_3$ ($z$=0.45 and 0.48)
could be changed over to its second ordered equivalent.
Not only in (Sm,Sr)MnO$_3$, the existence of
tricritical points in several other manganite systems have also been
observed, which will be discussed later.

On the theoretical modeling side, few works on FM-PM phase
transition of manganites have been reported
\cite{concen1,concen2,hysteresis,bustingorry} but none of them
characterize the nature of such phase transitions under the
presence of all three external parameters - pressure,
temperature and magnetic field. This is the principal focus of the present
study, namely to analyze phase transition in the three dimensional
phase volume. Our approach is based on a generalized version of
Landau theory, integrating all three variables/ parameters. We show
that both transition temperature and hysteresis width vary when
these parameters change. In a particular cases, there
exist tricritical points at which first-order FM-PM transition
changes to second-order. In order to illustrate the
presented picture, we analyze the experimental data of our previous
works \cite{prb08,prb09,apl1} on Sm$_{0.52}$Sr$_{0.48}$MnO$_3$
single crystal, an extensively studied material showing a
first-order FM transition.

\section{Theory}

In this section, we model the critical behavior related to the
FM-PM phase transition in the presence of magnetic field, pressure
and chemical substitution based on Landau theory. First we discuss
the effect of external magnetic field on the first-order FM phase
transition.

\subsection{Effect of external magnetic field on the FM-PM phase transition}

The magnetic order is described by the magnetization ${\bf{M}}=M
\hat{\bf{m}}$ such that $M=0$ defines the PM state and $M \neq 0$
represents the FM state. Expanding the free energy density around $M
= 0$, the magnetic free energy density in the presence of external
$H$ can be written as
\begin{equation}
F=F_0+\frac{1}{2}A\left(\frac {M}{M_S}\right)^2 -\frac
{1}{4}B\left(\frac {M}{M_S}\right)^4+\frac{1}{6} C\left(\frac
{M}{M_S}\right)^6- HM, \label{free1}
\end{equation}
where $F_0$ and $M_S$ are free energy density of PM phase and saturation
magnetization, respectively. In Eq.(\ref{free1}), the coefficient
$A$ can be assumed as $A = a(T-T^*) = a\tau$, where $a$ is a
positive constant and $T^*$ is the virtual transition temperature
\cite{skma}. The other coefficients $B$ and $C$ are assumed to be
temperature-independent and positive. In the absence of magnetic field ($H=0$),
the free energy density describes a first-order FM-PM phase
transition for $B>0$ while transition is second-order for $B<0$.
However, in presence of external field ($H\neq 0$), the transition
may become second-order even for $B>0$. In terms of scaled
magnetization $m = \frac{M}{M_S}$ and scaled magnetic field $h =
HM_S$, the free energy density can be written as
\begin{equation}
F=F_0+\frac 12Am^2 -\frac 14Bm^4+\frac 16 Cm^6-hm.
\label{free1a}
\end{equation}
The minimization of Eq.(\ref{free1a}) gives
\begin{equation}
Am-Bm^3+Cm^5-h=0, \label{cond1}
\end{equation}
from which one can obtain the differential equation for
susceptibility (defined as $\chi=\frac {1}{\chi_0}\frac {\partial
m}{\partial h}$ at $h$ = 0) as
\begin{equation}
(A-3Bm^2+5Cm^4)\chi-1/\chi_0=0 \label{cond1a}
\end{equation}
where $m$ is the spontaneous magnetization, which can be obtained
from Eq.(\ref{cond1}) in the absence of external magnetic field.
Eq.(\ref{cond1}) infers that $m=0$ can never be a solution of $h\neq
0$ phase, which means that an induced magnetization is observed in
the PM phase.

 {\color{black}{A key property of Eq. (\ref{cond1}) is the change from first- to
second-order FM phase transition at the critical point,
characterized by $h_C$, $\tau_C$ and $m_C$, where $h_C$, $\tau_C$
and $m_C$ are the critical values of magnetic field, temperature and
magnetization, respectively. The critical point is
obtained from the following condition
$F^{\prime}=F^{\prime\prime}=F^{\prime\prime\prime}=0$; this leads to 
\begin{equation}
h_C = H_CM_S = \frac {6B^2}{25C}\left(\frac {3B}{10C}\right)^{1/2};
~~~~~ \tau_C = T_C-T^* = \frac {9B^2}{20aC}; ~~~~~ m_C =
\frac{M_C}{M_S} = \left(\frac
{3B}{10C}\right)^{1/2}.\label{critical1}
\end{equation}
The complete solution of Eq. (\ref{cond1}) shows the variation of $m(\tau)$ with the temperature 
difference $\tau=T- T^{*}$ for
different $h$ values is shown in Fig. 1a, for which we have used the parameter values $a=1.0$, $B=3.0$, $C=8.0$. For $h$ = 0, $m$
drops very sharply (discontinuity in $m$ is around 0.43) at $\tau$ =
0.281. With increasing field strength, the magnitude of the magnetization jump (discontinuity) starts to decrease and eventually vanishes at the critical magnetic field $h_C \approx$ 0.09 at which zero-field FM-PM
transition changes from first- to second-order. This critical line differentiating the first and second order phase transition is identified by the dotted plot.}}

\begin{figure}[ht!]
\centering
    \begin{subfigure}[t]{0.48\textwidth}
    \centering
\includegraphics[height=0.42\textheight,width=1.1\linewidth]{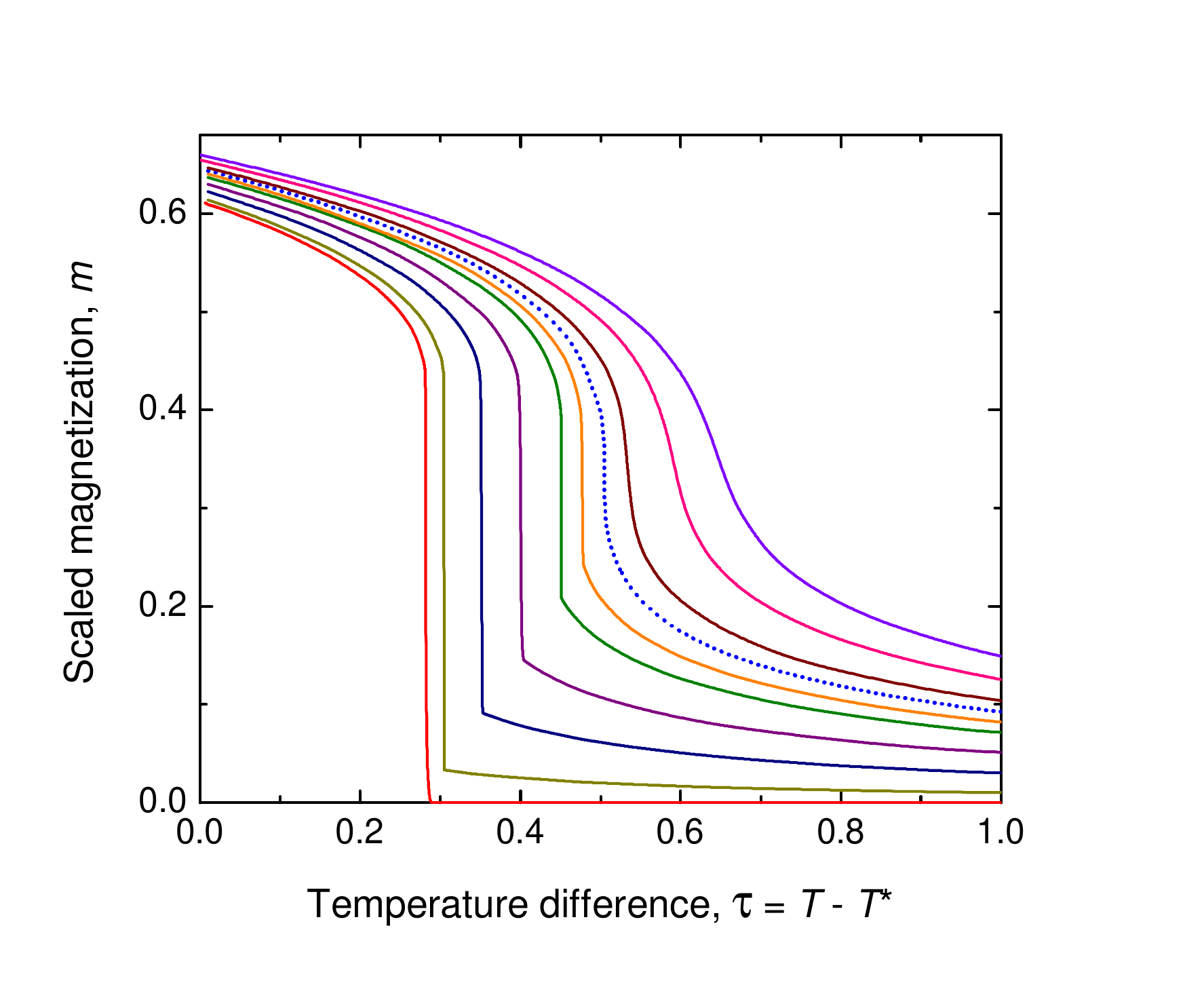}
        \caption{{\color{black}{Scaled magnetization, $m$ plotted against temperature difference, $\tau$ = $T - T^{*}$ for different scaled magnetic field values $h$ = 0 (extreme left curve), 0.01, 0.03, 0.05, 0.07, 0.08, 0.09, 0.1, 0.12 and 0.14 (extreme right curve). The dotted line corresponds to $h$ = 0.09 $\approx$ $h_C$. }}}
        \label{fig1a}
    \end{subfigure}%
 %   \hfill
    \quad
  \begin{subfigure}[t]{0.48\textwidth}
  \centering
\includegraphics[height=0.42\textheight,width=1.1\linewidth]{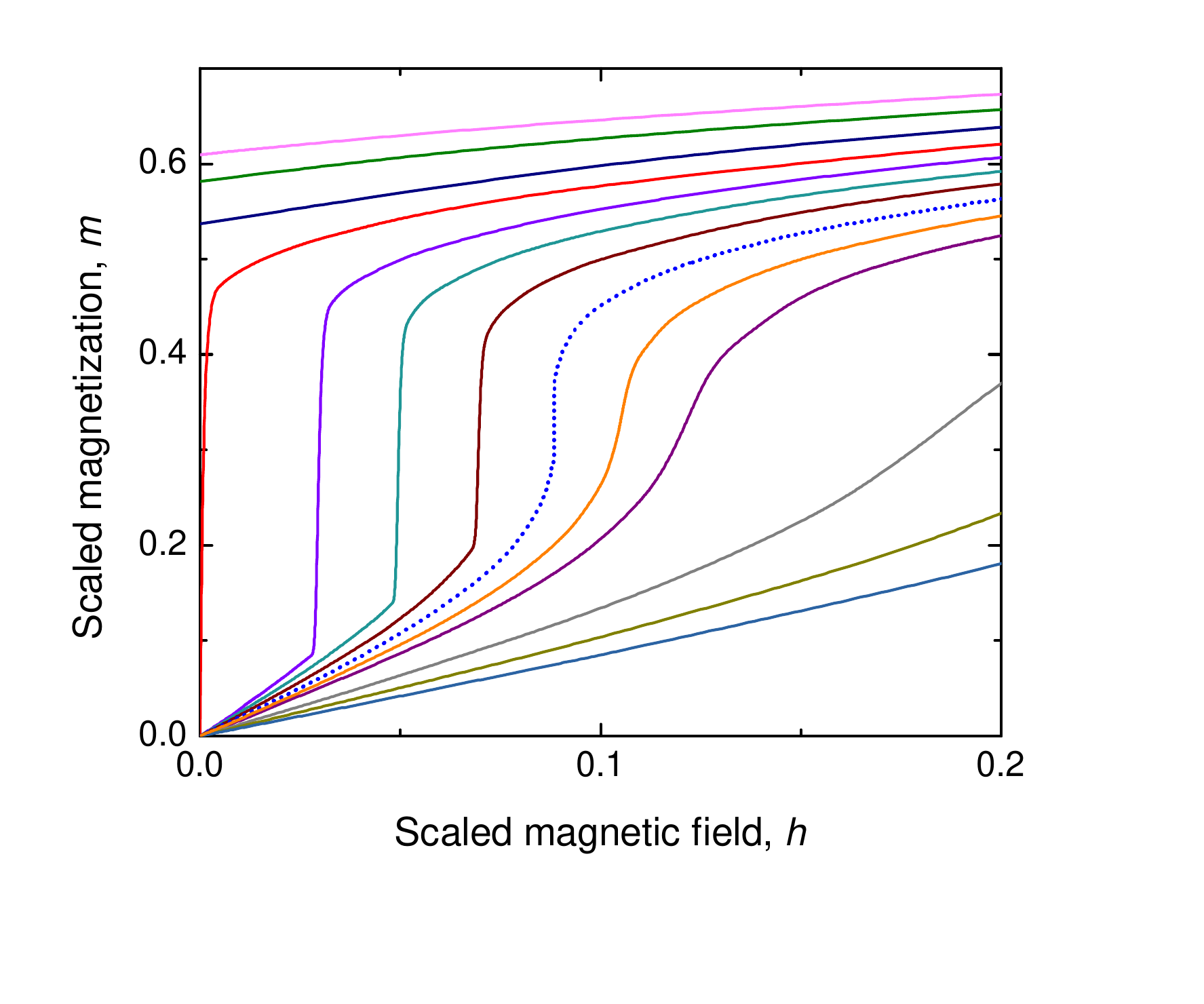}
        \caption{{\color{black}{$m(h)$ isotherms for different $\tau$. From top to bottom, each curve corresponds to $\tau$ = 0.01, 0.1, 0.2, 0.28, 0.35, 0.4, 0.45, 0.5, 0.55, 0.6, 0.8, 1.0 and 1.2. The dotted line corresponds to $\tau$ = 0.5 $\approx$ $\tau_C$. }}}
        \label{fig1b}
    \end{subfigure}
 %   \hfill
    \caption{(Color online) Variation of scaled magnetization $m$ against scaled magnetic field $h$ and temperature difference $\tau$ from the theoretical model. All the plots have been drawn for the parameter values $a$ = 1.0, $B$ = 3.0 and $C$ = 8.0.}
\end{figure}

Under the influence of $h$ ($h < h_C$), there is a shift in the
first-order FM transition temperature, {\color{black}{ which can be
calculated from the conditions $F-F_0=0$ and $\frac {\partial
F}{\partial m}=0$, as}}
\begin{equation}
T_{FM-PM}(h)=T^*+\frac {3B^2}{16aC}+\frac {10h}{4a}\left(\frac
{4C}{3B}\right)^{1/2}. \label{temp1}
\end{equation}
{\color{black}{The evolution of magnetization $m$ [according to Eq.(\ref{cond1})]
with magnetic field $h$ for different temperature values $\tau$ is
shown in Fig.~\ref{fig1b}. For $\tau_{FM-PM}(0) < \tau < \tau_C$,
magnetization isotherms exhibit field-induced characteristic jump
from low- to high-magnetic states. With increasing temperature, the
sharpness of the jump decreases and no such jump exists above
$\tau_C \approx$ 0.5. The values of $h_C$ and $\tau_C$ as observed
from Figures~\ref{fig1a} and \ref{fig1b} are consistent with Eq. (\ref{critical1}). Here
we would like to mention that the chosen values of the parameters
$a$, $B$ and $C$ are not unique, different values of them will
result different $h_C$ and $\tau_C$, but the nature of the phase
diagram remains unaltered.}}

%\captionsetup[subfigure]{labelfont=bf,textfont=normalfont,singlelinecheck=off,justification=raggedright}

Now we present our previously studied experimental
results analyzed on the basis of the theory discussed in the present
article. The studied system was Sm$_{0.52}$Sr$_{0.48}$MnO$_3$ (SSMO)
single crystal \cite{prb08,prb09}, which was prepared by floating
zone image furnace in oxygen atmosphere. The quality of the crystal
was carefully checked by various techniques such as x-ray powder
diffraction, Laue diffraction, electron dispersive x-ray analysis,
ac susceptibility, scanning electron microscope etc. The
magnetization measurements were performed in a superconducting
quantum interference device magnetometer in fields up to 7T using
five-scan averaging.
\begin{figure*}[ht!]
\centering
    \begin{subfigure}[t]{0.46\textwidth}
    \centering
\includegraphics[height=0.42\textheight,width=1.1\linewidth]{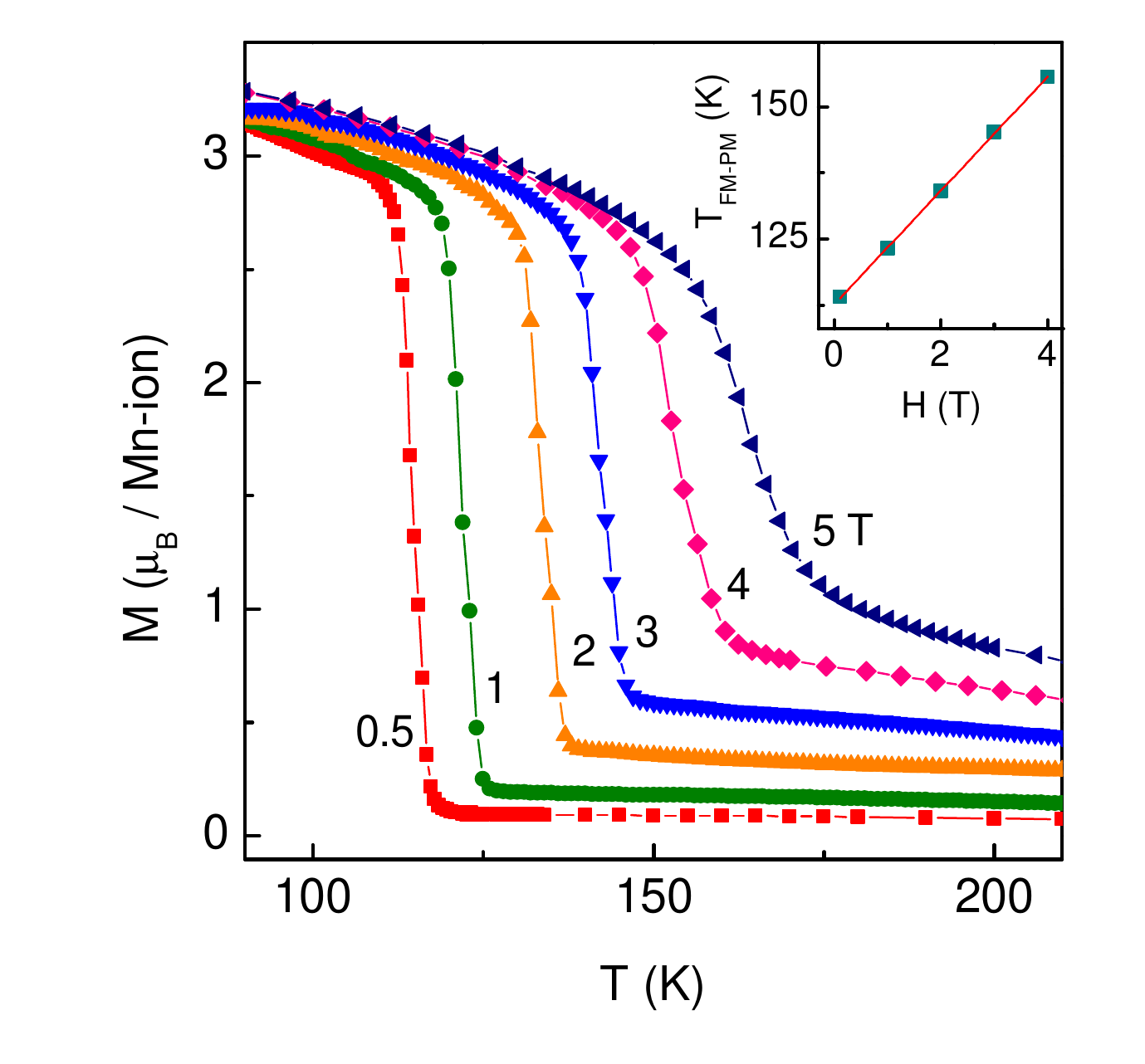}
        \caption{$M(T)$ curves of a SSMO crystal for a range of $H$ values. The inset shows $H$-dependence of ferromagnetic-paramagnetic transition
temperature $T_{\text{FM-PM}}$. The experimental data
points (symbol) are fitted (line) from a solution of {\color{black}{Eq. (2.6)}}.}
        \label{ssmomt}
    \end{subfigure}%
    \hfill
  \begin{subfigure}[t]{0.46\textwidth}
  \centering
%    \begin{subfigure}[b]{0.495\textwidth}
%        \centering
%        \includegraphics[height=8cm,width=9cm]{magT.pdf}
\includegraphics[height=0.42\textheight,width=1.1\linewidth]{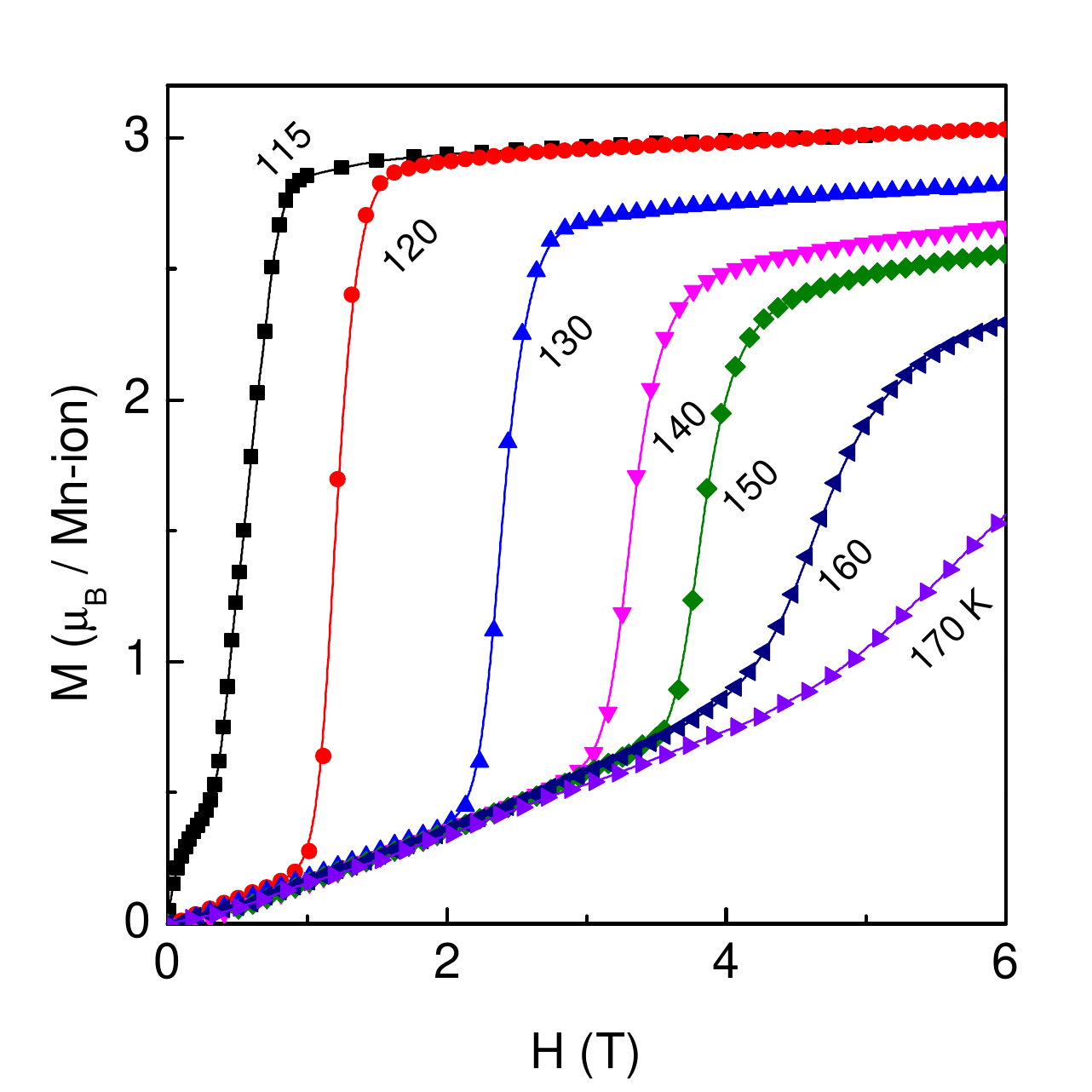}
        \caption{$M(H)$ isotherms of an SSMO crystal demonstrating the dependence of magnetization
($M$) on Magnetic field ($H$) of a Sm$_{0.52}$Sr$_{0.48}$MnO$_3$ (SSMO) single crystal for different temperatures ($T$).}
        \label{ssmomh}
    \end{subfigure}
 %   \hfill
    \caption{(Color online) Variation of magnetization (M) against magnetic field (H) and temperature (T) for a Sm$_{0.52}$Sr$_{0.48}$MnO$_3$ (SSMO) single crystal.}
\end{figure*}

%\begin{center}
%\begin{figure}[tbp]
%\includegraphics[height=16.0cm,width=12cm]{ssmom.pdf}
%\caption{(a) Temperature ($T$) dependence of magnetization ($M$) of
%Sm$_{0.52}$Sr$_{0.48}$MnO$_3$ (SSMO) single crystal for different
%magnetic field ($H$). Inset shows $H$ dependence of
%ferromagnetic-paramagnetic transition temperature ($T_{FM-PM}$). The
%experimental data points (symbol) are fitted (line) with Eq.
%(\ref{temp1}). (b) $M(H)$ isotherms of SSMO crystal. \label{ssmo}}
%\end{figure}
%\end{center}

Figure~\ref{ssmomt} shows $M(T)$ curves of SSMO crystal for a range of values of the
magnetic field $H$. In the small field regime, the sharp FM-PM
transition indicates that the transition is first-order in nature,
which gets weakened  with increasing $H$ as clearly reflected by the
suppression of the magnetization change associated with the
transition. Fig.~\ref{ssmomt} also depicts the nature of increase of
the FM-PM transition temperature with increasing external field;
here we have fitted the data displayed in this figure using Eq. (\ref{temp1}), specifically using the functional forms
$\left(T^{*}+\frac{3B^2}{16aC}\right)$ = 112.7 $\pm$ 0.2 K and
$\left(\frac{10M_S}{4a}\right)\sqrt{\frac{4C}{3B}}$ = 10.74 $\pm$
0.08 K/T as fitting parameters. In a related system
Sm$_{0.55}$Sr$_{0.45}$MnO$_3$ \cite{demko}, where Sr concentration
differs slightly, FM to PM phase transition is first-order with
$T_{FM-PM}(H=0) \sim$135 K. With the application of external field,
the transition is shifted to higher temperatures at an average rate
of d$T_{FM-PM}$/d$H$ = 9 K/T. Simultaneously, the first-order nature
of the transition weakens and above a critical point ($H_C \approx$
3.75 T, $T_C \approx$ 165.4 K), the transition becomes second-order.
{\color{black}{The field-induced change in the character of the FM-PM
transition is observed in other manganites also such as
Sm$_{0.55}$(Sr$_{0.5}$Ca$_{0.5}$)$_{0.45}$MnO$_3$ ($H_C \approx$
11.5 T, $T_C \approx$ 164 K) \cite{sscmo},
Eu$_{0.55}$Sr$_{0.45}$MnO$_3$ ($H_C \approx$ 7.4 T) \cite{demko} and
La$_{1-x}$Ca$_{x}$MnO$_3$ with $x$ = 0.25 ($H_C \approx$ 4 T)
\cite{ultrasonic} and 0.3 ($H_C \approx$ 6.5 T) \cite{jmmm17}. 

The evolution of $M$ with $H$ for different temperatures is shown in
Fig.~\ref{ssmomh}. It is clear that $M(H)$ isotherms are not
conventional Brillouin-like, rather they exhibit field-induced
metamagnetic transition. With increasing $T$, the sharpness of the
jump decreases. From a comparison with observations detailed in the
next section, it is clear that the temperature and field dependence
of magnetization observed in SSMO are qualitatively similar to the
magnetization behavior, calculated from free energy density
[Eq.(\ref{cond1})], as shown in Figures~\ref{fig1a} and
\ref{fig1b}. For SSMO, the experimentally observed critical point at
which zero-field first-order FM transition becomes second-order is
given by $H_C \approx 4 ~ T; ~~ M_C \approx 0.88 ~
\mu_B/\text{Mn-ion}; ~~T_C \approx 160 ~ K.$}} \\

\subsection{Effect of external pressure on the FM-PM phase transition}

\begin{center}
\begin{figure}[tbp]
\includegraphics[height=12.0cm,width=14.0cm,angle=0]{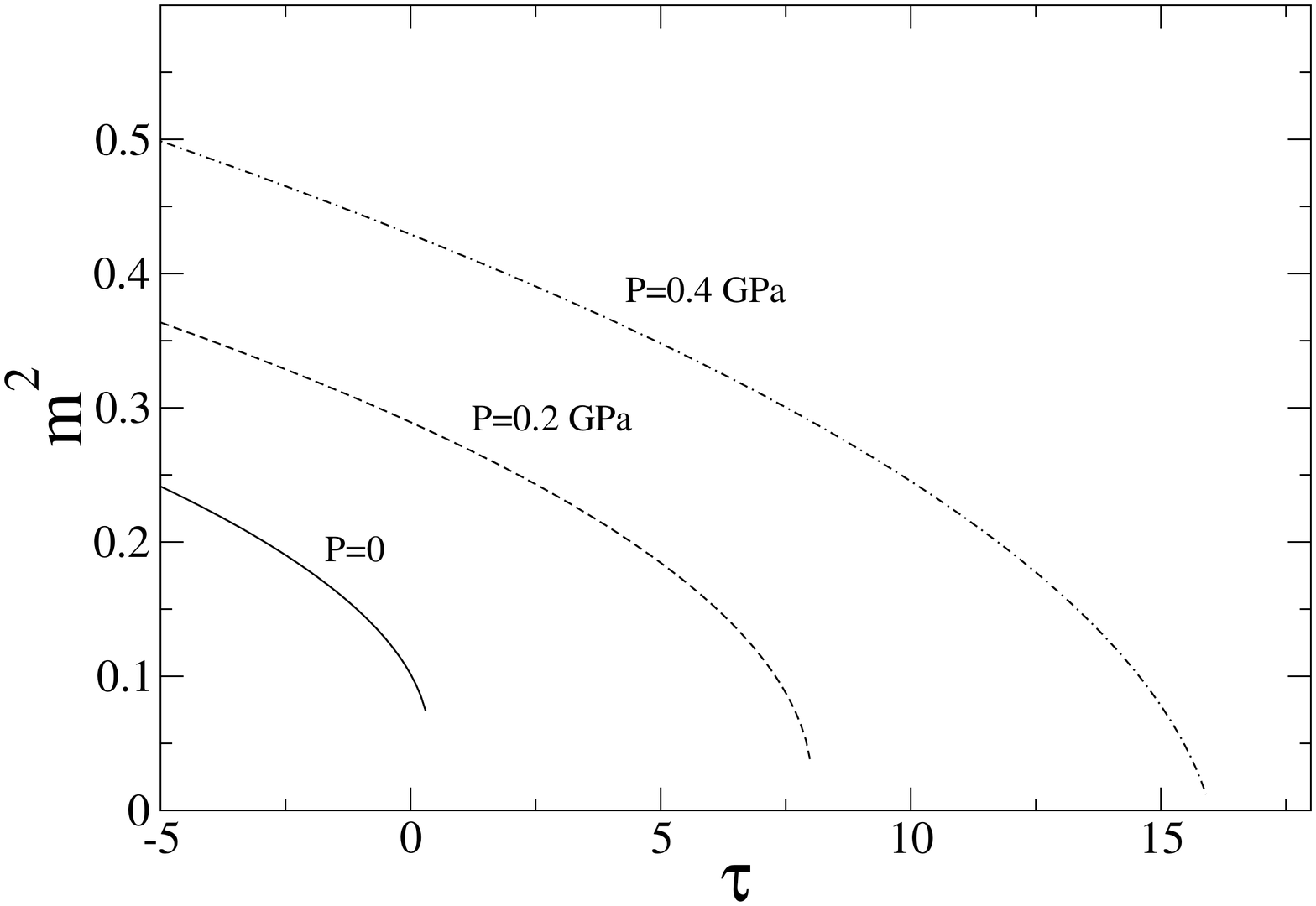}
\caption{{\color{black}{(Color online) $m^2$ vs. $\tau$ for different pressure. The plots have been
evaluated for the parameter values $a=0.1$, $B=0.5$, $C=0.8$,
$\eta_1=2.0$, $\delta_1=0.7$ and $u=0.5$.}}}
\label{magnetization}
\end{figure}
\end{center}

To study the pressure dependent change in FM-PM transition, we
consider a strong coupling between magnetic order parameter and  lattice
strain. So the free energy density in the presence of external
pressure can be written as
\begin{eqnarray}
F&=&F_0+ \frac 12A\left(\frac {M}{M_S}\right)^2 -\frac
14B\left(\frac {M}{M_S}\right)^4+\frac 16 C\left(\frac
{M}{M_S}\right)^6+\frac 12u_{ijkl}\epsilon_{ij} \epsilon_{jk}
\nonumber \\ &&-\frac 12\eta_1\epsilon_{ii}\left(\frac
{M}{M_S}\right)^2 +\frac 12 \delta_1\epsilon_{ii}^2\left(\frac
{M}{M_S}\right)^2-P_{ij}\epsilon_{ij} \label{free2}
\end{eqnarray}
{\color{black}{ The cross-coupling terms $\eta_1$ and $\delta_1$ are assumed to be
positive and they characterize the coupling strength between the
magnetic order parameter and strain tensor $\epsilon$. The positive signs of $\eta_1$ and $\delta_1$ ensure the increase of
magnetization and FM transition temperature with the increase of
pressure [see Eq. (\ref{magne1})]}}. The last term
of Eq. (\ref{free2}) represents the coupling between pressure and
elastic strain. In terms of scaled magnetization $m$, $F$ can be
rewritten as
\begin{equation}
F=F_0+\frac 12Am^2 -\frac 14Bm^4+\frac 16 Cm^6+\frac 12u\epsilon^2
-\frac 12\eta_1\epsilon m^2+\frac 12
\delta_1\epsilon^2m^2-P\epsilon, \label{free3}
\end{equation}
{\color{black}{Now the minimization of Eq. (\ref{free3}) with respect to
$\epsilon$ we get}}
\begin{equation}
\epsilon=\frac {P+\eta^* m^2+\delta^*m^4}{u}
\label{strain}
\end{equation}
with $\eta^*=\frac 12 \eta_1-\frac {\delta_1}{u}P$ and
$\delta^*=-\frac {\delta_1\eta_1}{2u}$. Equation (\ref{strain})
shows that strain changes with temperature and pressure since $m$
changes with temperature. Elimination of $\epsilon$ in Eq.
(\ref{free3}) yields
\begin{equation}
F=F_0^*+\frac 12A^{*}m^2-\frac 14
B^{*}m^4+\frac 16C^{*}m^6
\label{free4}
\end{equation}
where $F_0^*=F_0-\frac {P^2}{2u}$ and the renormalized coefficients
are given by
\begin{equation}
A^*=A-\frac {\eta_1 P}{u}+O(P^2), ~~ B^*=B+\frac
{\eta_1^2}{2u}-\frac {2\delta_1\eta_1}{u}P+O(P^2), ~~ C^*=C+\frac
{5\delta_1\eta_1^2}{4u^2}- \frac {5\delta_1^2\eta_1}{2u^3}P+O(P^2).
\label{coeff1}
\end{equation}
{\color{black}{The value of the magnetization in the ferromagnetic
state can be calculated after the minimization of Eq. (\ref{free4})
and can be expressed as}}
\begin{equation}
m^2=\frac
{B^*}{2C^*}\left[1+\left[1-\frac{4C^*}{B^{*2}}\left(a(T-T^*)-\frac
{\eta_1P}{u}\right)\right]^{1/2}\right], \label{magne1}
\end{equation}
{\color{black}{Eq.~(\ref{magne1}) shows that the magnetization increases with
increase of pressure which agrees with experiments
\cite{prb09,prl09}. In the FM phase, $T-T^*$ is negative and hence
the coupling constant $\eta_1$ should be positive to ensure the
increase of magnetization with increase of external pressure.
Moreover, the form of the free energy density as shown in Eq.
(\ref{free4}) clearly shows that the jump of the order parameter
$m_{FM-PM}=(3B^*/4C^*)^{1/2}$ decreases with increase of pressure.
To show more clearly the variation of the magnetization with
temperature as well as pressure in the FM phase, we have plotted
$m^2$ as a function of temperature for different pressure (Fig.
\ref{magnetization}). This is done for a set of phenomenological
parameters for which the FM-PM transition is possible. From Fig.
\ref{magnetization}, one can see that with increasing pressure, both
magnetization and FM transition temperature increase, whereas the
jump of the order parameter at the transition point diminishes,
indicating the closeness of second order character of the FM-PM
transition.}}

{\color{black}{The pressure dependent susceptibility can be calculated by adding a
term $-HM$ in the free energy expansion Eq.~(\ref{free4}) and then
the differential equation for susceptibility can be written as
\begin{equation}
(A^*-3B^*m^2-5C^*m^4)\chi-1/\chi_0=0, \label{cond1b}
\end{equation}
where the magnetization $m$ can be obtained from Eq. (\ref{magne1})
in the absence of the external magnetic field.}}
The pressure dependent first order FM-PM phase transition can be
{\color{black}{calculated from the condition
$F-F_0=0$ and $\frac {\partial F}{\partial m}=0$ as}}
\begin{equation}
T_{FM-PM}(P)=T^*+\frac {\eta_1P}{au}+\frac {3B^{*2}}{16aC^*}
\label{temp3}
\end{equation}
After simplification, Eq. (\ref{temp3}) can be rewritten as
\begin{equation}
T_{FM-PM}(P)=T^*+\frac {w}{16aC}+v_1P+O(P^2) \label{temp4}
\end{equation}
where $w=\left(B+\frac
{\eta_1^2}{2u}\right)^2\left(1-\frac{5\delta_1\eta_1^2}
{4u^2C}\right)$ and $v_1 = \frac
{\eta_1}{au}+\frac{5\delta_1^2\eta_1}{2u^3C}\left(1+\frac
{\eta_1^2}{2uB}\right)^2-\frac
{4\delta_1\eta_1}{u}\left(1-\frac{5\delta_1\eta_1^2}{4u^2C}\right)\left(1+\frac
{\eta_1^2}{2uB}\right)$.

{\color{black}{The spread of thermal hysteresis around the FM-PM phase transition
point can be calculated from the condition $F^{\prime}=F^{\prime
\prime}=0$ and is expressed as}}
\begin{equation}
\triangle T_{FM-PM}(P)=T^{**}-T_1^*=\frac {B^{*2}}{4aC^*}
\label{hyste1}
\end{equation}
where the supercooling temperature $T_1^*=T^*+\frac {\eta_1 P}{au}$. $T^{**}$ is the temperature of the superheated
ferromagnetic phase.
Equation (\ref{hyste1}) can be rewritten as
\begin{equation}
\triangle T_{FM-PM}(P)=\frac {w_1}{4aC}+v_2P \label{hyste2}
\end{equation}
where $w_1=B^2\left(1-\frac{5\delta_1\eta_1^2} {4u^2C}\right)$ and
$v_2=-\frac{5\delta_1^2\eta_1}{32u^3C}\left(B+\frac {\eta_1^2}{2u}\right)^2+\frac {\delta_1\eta_1}{4u}\left(1-\frac{5\delta_1\eta_1^2}{4u^2C}\right)\left(B+\frac {\eta_1^2}{2u}\right)$.

{\color{black}{From Eq.~(\ref{coeff1}), it is clear that as pressure changes the
renormalized coefficients $A^*$ and $B^*$ change and hence order of
the FM transition also changes. For weak coupling and the lower
value of pressure, $B^*>0$, \emph{i}.\emph{e}., $B> \frac
{2\delta_1\eta_1}{u}P - \frac {\eta_1^2}{2u}$, indicating that the
FM-PM phase transition is first order, where both FM and PM phases
coexist, \emph{i.e.}, a thermal hysteresis appears [see Eq.
(\ref{hyste1})]. As pressure increases, $B^*$ starts to decrease
and for strong coupling and high value of $P$, $B^*<0$ i.e. $B < \frac {2\delta_1\eta_1}{u}P - \frac {\eta_1^2}{2u}$, then a second order
transition occurs. Thus at critical pressure $P_C$, thermal
hysteresis vanishes and the first-order FM-PM transition becomes
second-order in nature. For a particular value of the pressure, $B^*
= 0$, then the first-order FM transition crosses over to the second order
transition i.e. a tricritical point is obtained. Hence, a
tricritical point can be achieved by varying external pressure $P$ only}}. The quantitative nature of the pressure/doping dependence on magnetization explained through Eqs. (2.13-2.17) are structurally reminiscent of Eqs. (2.3-2.6), and hence will show qualitative similarity with Figs. 1 and 2 discussed earlier.
\begin{center}
\begin{figure}[tbp]
\includegraphics[height=10.0cm,width=12.0cm]{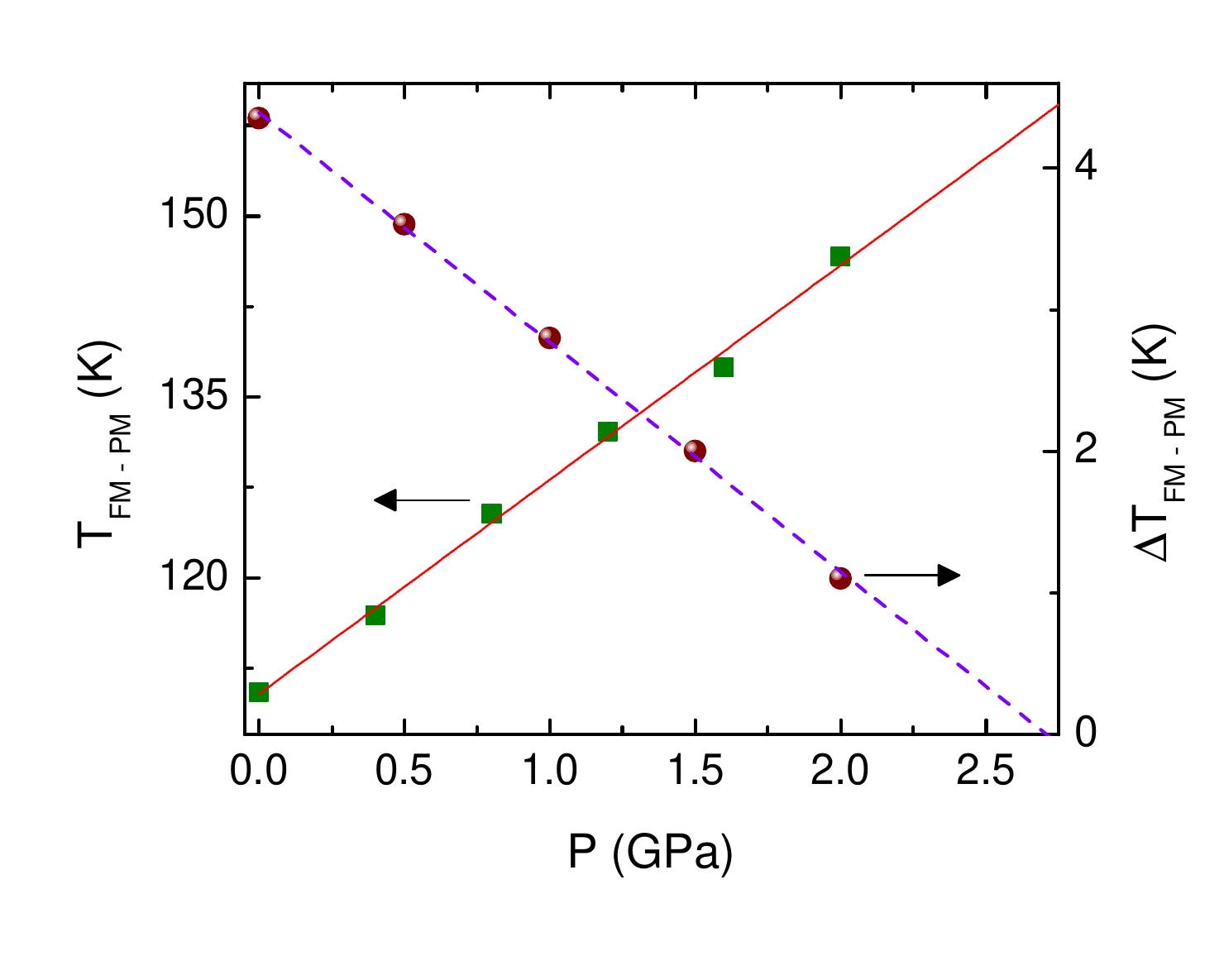}
\caption{(Color online) Pressure ($P$) dependence of FM transition temperature
($T_{FM-PM}$) and thermal hysteresis width ($\Delta T_{FM-PM}$) of
SSMO single crystal. The symbols are experimentally measured data
points and the solid and dashed lines are the best fit of Eqs.
(\ref{temp4}) and (\ref{hyste2}), respectively. \label{fig5}}
\end{figure}
\end{center}
In order to justify our proposed theory, we present
our previous experimental work on SSMO single crystal
\cite{mydeen,prb09}. The effect of external pressure (up to 2 GPa)
on the nature of FM to PM phase transition has been studied. With
increasing pressure, $T_{FM-PM}$ increases while the width of
thermal hysteresis reduces as shown in Fig. (\ref{fig5}). We fit
these experimentally measured data points according to Eqs.
(\ref{temp4}) and (\ref{hyste2}) which clearly points to the strong
agreement of our theoretical model based results with real experimental data. For this, we have used the fit
parameters $\left(T^* + \frac{w}{16aC}\right)$ = 110.4 $\pm$ 0.7 K,
$v_1$ = 17.8 $\pm$ 0.5 K/GPa, $\frac{w_1}{4aC}$ = 4.39 $\pm$ 0.04 K
and $v_2$ = -1.62 $\pm$ 0.03 K/GPa. For SSMO, the value of critical
pressure where the zero-field transition becomes second-order is
$P_C \approx 2.7$ GPa. In Sm$_{0.55}$Sr$_{0.45}$MnO$_3$
\cite{demko}, the application of pressure increases $T_{FM-PM}$
linearly at the rate of $\sim$20 K/GPa, while $\Delta T_{FM-PM}$
narrows down. The critical pressure where the transition changes its
character is estimated to be $P_C \approx$ 3.2 GPa. \\

\subsection{Effect of chemical substitution on the FM-PM phase transition}

Let us now consider the effect of chemical substitution on the FM-PM
phase transition. In the case of binary mixture of impurity, the
free energy must be expressed in terms of the order parameter and
impurity concentration $x$. The simplest way to take into account
the effect of impurity is to introduce the impurity-magnetization
coupling terms in the free energy expression which becomes
\begin{equation}
F=F_0+\frac 12A\left(\frac {M}{M_S}\right)^2 -\frac 14B\left(\frac
{M}{M_S}\right)^4+\frac 16 C\left(\frac {M}{M_S}\right)^6 -\frac 12
D\left(\frac {M}{M_S}\right)^2x+\frac 12E \left(\frac
{M}{M_S}\right)^2x^2+\frac 12Gx^2. \label{free5}
\end{equation}
{\color{black}{where $A=a(T-T^*)$}}. The term $\frac 12Gx^2$ is the
free energy density of the impurity solute. $D$ and $E$ are the
coupling constants. {\color{black}{All the coefficients $B$, $C$, $D$, $G$ and $E$ are assumed to be
positive to ensure the increase of transition temperature [see
Eq.(\ref{temp6})] with $x$. Since the order of FM to PM phase
transition depends on the sign of the coefficient $B$. We assume $B$
changes with concentration and we set $B=b_0(x-x_0)$, where $x_0$ is
the equilibrium value of concentration and $b_0$ is a positive
constant \cite{pkm}.}}

Taking the partial derivate of Eq. (\ref{free5}) with respect to
$x$, we get
\begin{equation}
\frac {\partial F}{\partial x }=Gx-\frac 12 Dm^2+Em^2x \equiv \mu,
\label{cond2a}
\end{equation}
where $m = \frac{M}{M_S}$ and $\mu$ is the quantity
thermodynamically conjugate to $x$. Simplifying Eq. (\ref{cond2a}),
we get
\begin{equation}
x = \frac 1G\left[\mu \left(1-\frac {Em^2}{G}\right)+\frac 12Dm^2-
\frac 12 \frac {ED}{G}m^4\right]. \label{cond2b}
\end{equation}
Applying Legendre transformation, we have
\begin{eqnarray}
F(m,\mu,T )&=&F(m,x,T)-\mu x \nonumber \\
&&=F_0^{**}+\frac 12A^{**}m^2-\frac 14B^{**}m^4+\frac 16C^{**}m^6
\label{free6},
\end{eqnarray}
where $F_0^{**}=F_0-\frac {\mu^2}{2G}$ and the renormalized
coefficients are given by

\begin{equation}
A^{**}=A-\frac {\mu D}{G}+\frac {\mu^2 E}{G^2}; ~~~ B^{**}=B+\frac
{7D^2}{2G}-\frac {\mu ED}{G^2}+\frac {\mu^2 E^2}{G^3}; ~~~
C^{**}=C+\frac {21 ED^2}{4G^2}+\frac {3\mu^2 E^3}{G^4}.
\label{coeff4}
\end{equation}

{\color{black}{It is clear from Eq. (\ref{coeff4}) that the
consideration of couplings between magnetization $M$ and impurity
concentration $x$ leads to the renormalization of the coefficients
$A$, $B$ and $C$. The coefficient $B$ changes with $x$, which means
that the order of FM-PM transition can also change with impurity
concentration. For weak coupling constants $E$ and $D$ and low value
of $x$, $B^{**}>0$ and $C^{**}>0$, \emph{i.e.}, FM-PM transition is
first-order in nature. Then the concentration dependence of FM-PM transition
temperature can be calculated following the procedure as in
Eq.~(\ref{temp3}) to get
\begin{equation}
T_{FM-PM}(x)=T^*+\frac {3B^2}{16aC}+\frac {Dx}{a}-\frac {Ex^2}{a}.
\label{temp6}
\end{equation}
Similar to the procedure adapted in Eq. (\ref{hyste1}), the width of
thermal hysteresis is given by
\begin{equation}
\triangle T_{FM-PM}(x)=\frac {3B^{2}}{4aC} \label{hyste3}
\end{equation}
Equation (\ref{temp6}) shows that first order FM-PM transition
temperatures increases with the increase of concentration. This
shows that the coupling constants $E$ and $D$ should be positive.

Now for strong coupling constants $E$ and $D$ and higher value of
concentration $x$, $B^{**}<0$, the transition is second order. For a
particular value of the concentration $x=x_{tcp}$, $B^{**}=0$, the
first order transition goes to a second order transition. So, there
is a crossover from first to second order transition and a
tricritical point appears}}.

\begin{center}
\begin{figure}[tbp]
\includegraphics[height=10.0cm,width=12.0cm]{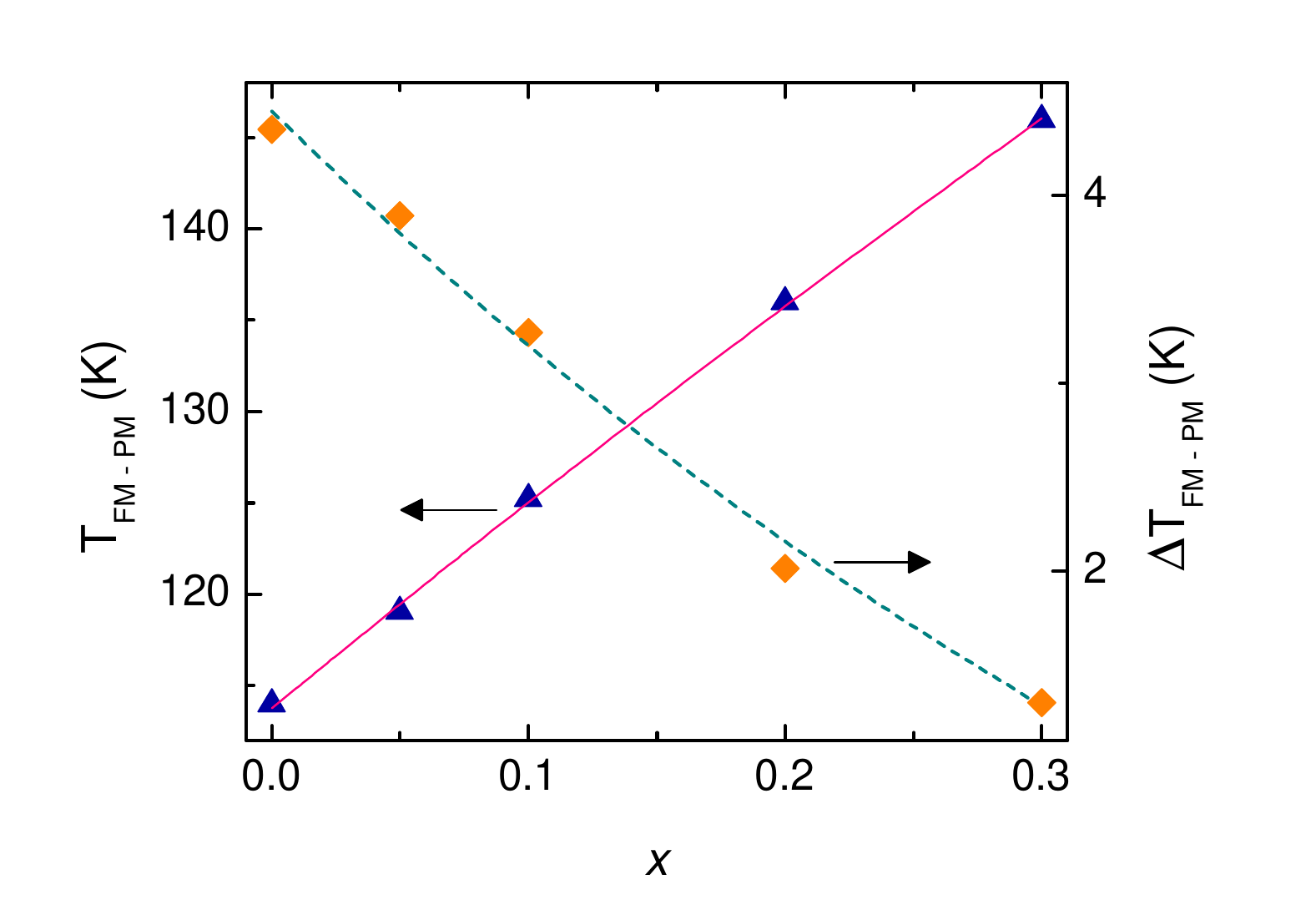}
\caption{(Color online) Ferromagnetic-paramagnetic transition temperature
($T_{FM-PM}$) and thermal hysteresis width ($\Delta T_{FM-PM}$) of
(Sm$_{1-x}$Nd$_x$)$_{0.52}$Sr$_{0.48}$MnO$_3$ single crystals as a
function of Nd concentration ($x$). The symbols represent
experimental data points while the solid and dotted lines are the
best fit of Eqs. (\ref{temp6}) and (\ref{hyste3}), respectively.
\label{fig6}}
\end{figure}
\end{center}

From the condition $A^{**}=0$, the second order transition can be
expressed as
\begin{equation}
T_C=T^*+\frac {Dx}{a}-\frac {Ex^2}{a} \label{temp7}
\end{equation} \\
In our previous experimental study \cite{prb09}, we have observed
the doping dependence change in the order of FM-PM transition. The
studied system was (Sm$_{1-x}$Nd$_x$)$_{0.52}$Sr$_{0.48}$MnO$_3$
single crystals with $0 \le x \le 0.3$. The magnetization data of
parent compound SSMO ($x$ = 0) shows a first-order FM to PM phase
transition at $T_{FM-PM} \approx$ 114 K along with thermal
hysteresis of width $\triangle T_{FM-PM} \approx$ 4.3 K. With the
substitution of Nd at Sm site, $T_{FM-PM}$ increases while
$\triangle T_{FM-PM}$ decreases, which are displayed in Fig.
\ref{fig6}. These experimental data (solid symbols) are fitted to
Eqs. (\ref{temp6}) and (\ref{hyste3}). The best-fit parameters are
given by $T^*$ = 109.3 $\pm$ 0.4 K, $\frac {B_0^2}{16aC}$ = 3.54
$\pm$ 0.07 K, $x_0$ = 0.64 $\pm$ 0.03, $D/a$ = 129.7 $\pm$ 0.6 K,
$E/a$ = 38.1 $\pm$ 0.6 K; the corresponding curves (solid lines) are
as shown in Fig. \ref{fig6}. The complete phase diagrams of
(Sm$_{1-x}$Nd$_x$)$_{0.55}$Sr$_{0.45}$MnO$_3$ single crystals have
been studied for $0 \le x \le 0.5$ by Demk\'{o} \emph{et al.}
\cite{demko} . For $x$ = 0, FM transition is first-order with
$T_{FM-PM} \approx$ 135 K and $\triangle T_{FM-PM} \approx$ 1.7 K.
With increasing Nd concentration, $T_{FM-PM}$ increases and
$\triangle T_{FM-PM}$ narrows down. Above a critical concentration
$x_C \approx$ 0.33, the hysteresis completely vanishes and the
first-order transition becomes second-order. Similar kind of
behavior has also observed in La$_{1-z}$Ca$_{z}$MnO$_3$, where
first-order ($z <$0.4) FM to PM phase transition changes to
second-order ($z >$0.4) with increasing Ca doping \cite{kim}.

\section{Conclusions}

We have discussed the FM-PM phase transition based on a
phenomenological Ginzburg-Landau theory. In the presence of magnetic field, first order
FM-PM phase transition can become second ordered at the critical
point. The effect of pressure on the FM-PM phase transition is to
increase the transition temperature. The theory predicts a second
order character of the FM-PM phase transition at high pressure. The
formation of two phase regions during the FM-PM phase transition, prompted by chemical
substitution, depends on the small
values of the coefficients $B$ and $C$. The different values of the
Landau coefficients indicate the change of the transition
temperatures and the second order character of the transition. In a
mixture, the FM-PM phase transition which is first order of the pure
form becomes a second ordered transition with the change of concentration. This
leads to a crossover from first to second order transition via a tricritical point. This study thus presents a theory based confirmation of
experimental results both at the qualitative and quantitative levels.

{\color{black}{As shown from comparison with other experimental results on manganites like 
Sm$_{0.55}$(Sr$_{0.5}$Ca$_{0.5}$)$_{0.45}$MnO$_3$, Sm$_{0.55}$Sr$_{0.45}$MnO$_3$, Sm$_{0.52}$Sr$_{0.48}$MnO$_3$, 
Sm$_{0.55}$(Sr$_{0.5}$Ca$_{0.5}$)$_{0.45}$MnO$_3$, Eu$_{0.55}$Sr$_{0.45}$MnO$_3$ and La$_{1-x}$Ca$_{x}$MnO$_3$, the variation of FM-PM transition temperature and hysteresis width with external/ internal perturbations conform to our model results.}}
Changes in the character of FM phase transition under the influence of external magnetic field, pressure and chemical substitution observed in several manganite systems \cite{kim, ultrasonic,mydeen,prb08,demko,prb09, prl09, sscmo,sr17, apl1, apl2} have all been satisfactorily explained within the remit of our theoretical model justifying
the occurrence of a tricritical point, a hitherto unexplained feature in most studies.

\section{Acknowledgment}
PS is sincerely grateful to Professor P Mandal for his many valuable
suggestions and for a critical reading of the manuscript. AKC
acknowledges partial financial support to SARI, Aston.

\end{document}